# Martensitic transformation in zirconia
# Part II: Martensite growth

Sylvain Deville, Gérard Guénin, Jérôme Chevalier

Associate Research Unit 5510, Materials Science Department, National Institute of Applied Science (GEMPPM-INSA), Bât B. Pascal, 20 av. A. Einstein, 69621 Villeurbanne Cedex, FRANCE

**Abstract**

Though the martensitic transformation in zirconia has been the object of a very large number of studies for the last decades, qualitative and quantitative observations of the formation and growth of relief induced by low temperature treatments has hardly ever been reported. In the first part of the study, we have demonstrated the excellent agreement between the atomic force microscopy quantitative observations and the outputs of the calculations derived from the phenomenological theory of martensitic transformation. The intermediate stages of transformation were nonetheless not considered. In this second part, the growth mechanisms of monoclinic phase resulting from the martensitic transformation in ceria-stabilized zirconia (10 mol% $CeO_2$) are investigated. Surface transformation is induced by ageing treatments in water vapor at 413 K. The observations are rationalized by the recent analysis proposed for the crystallographic ABC1 correspondence choice, where the $c_t$ axis transforms to the $c_m$ axis. Three growth modes are observed and interpreted in terms of transformation strains accommodation. Microcracks formation is observed, explaining grain pop-out where the crystallographic disorientation between two adjacent grains is the largest. The influence of grain boundary paths on the surface relief features is demonstrated. Overall, our results strongly support the non-existence of a critical grain size for low temperature transformation, confirmed by the classical thermodynamics theory applied to this particular case.

**Résumé**

Malgré les nombreuses études dédiées à la transformation martensitique de la zircone au cours des dernières décades, des observations qualitatives et quantitatives de la formation et la croissance du relief induit par la transformation au cours de traitements à basse température n'ont jamais été rapportées. Dans la première partie de cette étude, nous avons démontré l'excellente concordance entre les observations expérimentales quantitatives par microscopie à force atomique et les prédictions des calculs obtenus par l'application de la théorie phénoménologique de la transformation martensitique. Les étapes intermédiaires de la transformation n'étaient toutefois pas prises en considération. Dans cette deuxième partie, les mécanismes de croissance de la martensite (phase monoclinique) dans la zircone cériée (10 mol% $CeO_2$) sont étudiés. La transformation de surface est induite par des traitements en autoclave à basse température (413 K), en présence de vapeur d'eau. Ces observations sont interprétées par l'analyse récemment proposée pour la correspondance ABC1, où l'axe $c_t$ se convertit en l'axe $c_m$. Trois modes de croissance sont observés et interprétés en termes d'accommodation des déformations de transformation. La formation de microfissures est observée, menant à un arrachement de grains où la désorientation cristallographique entre les grains adjacents est la plus forte. L'influence du chemin des joints de grains sur les caractéristiques du relief de surface est démontrée. Nos résultats supportent finalement fortement la non-existence d'une taille critique de transformation à basse température, conclusion appuyée par la théorie thermodynamique classique appliquée à ce cas particulier.

Mots-clés : *zircone, transformation martensitique, microscopie à force atomique, microfissure*

## 1. Introduction

Martensitic transformations occur in a number of metal alloys but also minerals and ceramics like zirconia [1, 2]. In order to stabilize zirconia and retain it in its metastable tetragonal structure at ambient temperature, it is possible alloying it with various oxides, among which yttria ($Y_2O_3$) and ceria ($CeO_2$). If the metastable tetragonal structure is indeed retained at ambient temperature, the transformation can however occur during low temperature treatment in water vapour [3, 4]. This phenomenon is known as ageing, and has been widely investigated for its detrimental consequences on the long-term performance of zirconia components. Several conclusions were drawn from the experimental observations, performed mostly by TEM, SEM and XRD, among which the existence of a critical grain size was observed, size varying from 0.1 to 0.5 µm in the case of 3Y-TZP [5, 6]. This existence of a critical grain size was demonstrated by the application of the classical thermodynamics theory [7, 8]. Moreover, it was argued the transformation was very different in the case of Y-TZP and Ce-TZP, because of the respective trivalent and quadrivalent nature of the stabilizing specie, leading to a different oxygen vacancies concentration [4]. As a consequence, the ageing sensitivity of these two materials was expected to be very different. Finally, microcracking was observed in the surroundings of transformed grains, and related to the volume and shear components of the transformation. However, no direct observation and clear interpretation of the microcracks formation has been reported. The observation of martensite formed during ageing treatment is very appealing in regards of the lack of validation of the various theories, for several reasons. The transformation is occurring at surface so that it may be observed straightaway, and the influence of a free surface on the transformation can potentially be assessed. Moreover, the transformation is propagating step by step when performing several ageing treatments in autoclave, so that intermediate stages of the transformation should be observed. Then, the less stable parts (from an energetical point of view) of the surface are the first one to transform; the transformation is not triggered by the experimental observations, like grains transforming under the beam during TEM observations. Finally, the sample geometry has no effect on the transformation behaviour, as opposed to thin foils where the very low thickness induces some peculiar features. It should therefore be possible determining some of the factors affecting the stability.

Considering the scale at which the transformation is occurring, very few qualitative and quantitative reports of transformation induced relief can be found in the literature. Considering the dilatational (0.04) and shear (0.16) components of the transformation, this dearth of experimental observations is not surprising; modifications of the surface relief occurs below the micrometer range, typically from 10 to 100 nm for a zirconia polycrystal of typical grain size (0.5 to 3 µm). In absence of experimental validation of the relief features, the predictions of transformation induced relief rely on the validity and relevance of the outputs of the phenomenological theory [9, 10] of martensite transformation. Validation of these predictions requires precise quantitative measurements of martensite relief. The effect of free surface on the transformation local characteristics relies mostly so far on the outputs of calculations [11-13]. Preliminary reports [14-16] have nonetheless drawn the attention of the potentialities offered by atomic force microscopy, technique offering a vertical resolution below the nanometer range. AFM allows straightforward observations of transformed surfaces of bulk samples. The observation of partial transformation of the grains [16] already questioned the existence of a critical size for the transformation. The technique has just been applied [17] to investigate the martensitic relief at the end of the transformation, and compared with the outputs of the crystallographic theory of martensitic transformation. An excellent quantitative agreement was found between the experimental observations and theoretical predictions. In particular, the peculiar behaviour of correspondence ABC1 was demonstrated. For this correspondence, it was shown the characteristics of the habit planes and transformation matrix allow a complete accommodation of transformation strains by a surface uplift outside of the free surface. No residual stresses should be expected in that case. A particular interest of this analysis relies in the interpretation of martensitic variants arrangement in the volume of the material from the experimental observations of relief features at the surface.

On the other hand, if the martensitic phenomenological theory was successfully applied, it is worth remembering the theory does only describe the transformation in mathematical terms, and by no means in physical terms. No informations on the chemical mechanism of transformation are brought by the theory. If the spatial arrangement was well understood, in terms of transformation strain accommodation, the remaining question was thus: how are these arrangements formed? How do they grow and how fast?

In this study, this approach is applied to investigate and analyse the nucleation and growth of martensite in ceria stabilized zirconia and subsequent consequences like microcracking. It is shown how the combination of AFM observations and of the outputs of the phenomenological analysis can provide new insights on both the physical and the chemical mechanisms of the transformation, with particular attention being paid on the influence of the free surface on the variants growth modes. AFM observations were performed at different steps of the ageing treatment in order to follow locally the transformation at the surface of the sample with time.

## 2. Experimental methods

Ceria stabilized zirconia (Ce-TZP) materials were processed by classical powder processing route, using Zirconia Sales Ltd powders, with uniaxial pressing and sintering at 1823 K for two hours. Residual porosity was negligible. Samples were mirror polished with standard diamond based products.

The martensitic transformation was induced by a thermal treatment in water vapour autoclave. This kind of treatment is known to induce the tetragonal to monoclinic phase transformation in zirconia [3, 18, 19]. Hence, these treatments were conducted in autoclave at 413 K, in saturated water vapour atmosphere, with a 2 bar pressure, inducing phase transformation at the surface of the samples with time. Thermal treatment steps were bounded to the thermal activation of the transformation and the technical limits of the autoclave. These steps could have been reduced by several decades if a higher treatment temperature was chosen.

AFM experiments were carried out with a D3100 nanoscope from *Digital Instruments Inc.*, using oxide sharpened silicon nitride probes (*Nanosensor*, *CONT-R* model) in contact mode, with an average scanning speed of 10 µm.s$^{-1}$. Since the t-m phase transformation is accompanied by a large strain, surface relief is modified by the formation of monoclinic phase. The vertical resolution of AFM allows following very precisely the transformation induced relief.

# 3. Experimental Results

## 3.1 Variants growth modes

Three different modes of variants growth are experimentally observed for the correspondence ABC1, leading to a four-fold symmetry final arrangement. These three modes will be respectively referred to as *internal growth*, *external growth* and *needle growth*.

- Internal growth

This first mode of transformation is related to the partial transformation, reported in [17]. It was shown how obtaining a stack of four variants of correspondence ABC1 was possible and energetically stable, since all interfaces (inside and outside of the arrangement) between tetragonal and monoclinic phases were habit planes of the same type. All the transformation strain is accommodated by this configuration. Fig. 1a provides such an observation, with its evolution as a function of the ageing treatment time. Treatments steps of 20h at 413 K in autoclave were performed between each image. The progressive transformation of the inner tetragonal part of the arrangement is clearly observed. This behaviour is interpreted in Fig. 1b, with the progressive growth of variants, toward the inside of the grain. The potential future habit planes are symbolized by the dashed line, the transformation strain being constant along these planes. Hence, the variants can grow until the transformation is completed, without being restrained by any transformation induced stresses. The way the primary variants can grow to form the initial fourfold symmetry arrangement will be discussed later, in regards of the other variants growth mode. This mode will be called "internal growth", in regards of its peculiar features.

- External growth

This mode is the opposite of the first one. In this case (Fig. 2a), the transformation starts undoubtedly at the surface, with the formation of a fully accommodated set of opposed variants, of fairly small size. The following observations show the progressive growth of the variants, the trace of the junction plane being in a constant location. This behaviour is well understood considering the variants progressively grow into the volume of the sample, while the junction plane remains in the same position (Fig. 2b). All the transformation strain is again accommodated; the variants are at equilibrium during all the growth steps. More complex arrangements, like the so-called "L-arrangement" [17] can be obtained by this mode, as shown in the last micrographs of Fig. 2a. Since the variants grow in a direction opposed to the junction plane, this mode will be called "external growth".

- Needle growth

The last growth mode observed in these experiments corresponds to the transformation sequence observed in Fig. 3. Again, transformation is very likely nucleating at the surface. An isolated martensite needle is formed and grows in size.

As soon as it is formed, it begins to generate opposite stresses (back stresses) in the surrounding matrix, opposing continued transformation of the initial variant. Since no complementary variants that could accommodate the transformation strain are observed in the first steps, the back stresses are building up with as the variant thickens, until the formation of a second variant is triggered by the shear components of the back stresses. This second variant will again grow in size and trigger the formation of a third variant. With this last configuration, the added stresses are so high that the transformation completion of the grain cannot proceed anymore, until an external event can provide supplementary stresses to overcome the energy barrier opposed to transformation completion. The remaining untransformed part is transformed very rapidly. The grain goes from very partial transformation to transformation completion during the last treatment step.

Though no statistical analysis was performed on the different growth modes, it seems from our experimental observations the internal growth mode is the first mode activated, after an apparent initial incubation period of about 60 hrs, during which no surface transformation at all seemed to be observed. The external growth mode follows, and finally the isolated needle growth appears. Hence, the controlling factors of each growth mode must be different.

## 3.2 Variants growth kinetics

Using experimental observations like in Fig. 1a, 2a and 3, the size of each variant can be measured as a function of their age, so that the kinetic of variants growth are obtained for every growth mode. These kinetics are given respectively in Fig. 4, 5 and 6. The width is normalised to the width of the fully transformed variants for comparison between the three modes. Growth speeds in each case were measured before normalisation.

Differences are observed between the three modes, though similarities are also found. The simplest case is that of external growth (fig. 4). The width increase of the variants is proportional to the treatment time; the growth rate is thus constant. As far as the internal growth is concerned (fig. 5), transformation occurs in two stages. In the first one, the variants width increases rapidly, and then approaches a lower and nearly constant growth rate, until transformation completion. A more complex behaviour is observed needle growth mode (fig. 6); transformation occurs in three stages. The first two stages are similar to that of internal growth, with a progressive decrease of the growth kinetics. A third stage is finally observed with a very fast growth speed to complete the transformation.

The distinction between these three stages is based on the measurements of growth rate, given in Tab. 1, and can be rationalized by strain accommodation arguments. For the three modes, the *initial* growth rate is very similar (11nm/h at 140°C), and must be therefore related to a similar mechanism. It is therefore proposed to correspond to the *unconstrained* growth. At the very beginning of transformation, the magnitude of transformation induced stress due to non-accommodation (for needle growth) is so low that the variants are free to grow at a similar speed than in the case of a nearly perfect accommodation demonstrated for the two other modes (internal and external growth). However, as these back stresses are building up very rapidly in the regions of misfit as the variant thickens, the growth rate falls down very rapidly. This corresponds to the steady rate of growth (stage II), where the growth rate is smaller than 3 nm/h and continuously decreasing with ageing time. The growth rate of stage II of the internal growth is similarly low, but must be related to a different phenomenon. In this case, it has been shown the transformation strains were continuously accommodated during the transformation. No induced stresses will act to slow down the transformation. Considering the proposed analysis (Fig. 1b), the progressive diminution of the size of the inner untransformed part is likely to decrease the driving force for transformation, so that the overall growth rate decreases. By using the model proposed for the volumic arrangement of the variants [17], it was possible calculating the remaining volume of untransformed tetragonal phase by using the measurements of the apparent traces of habit planes at surface. The estimation of the remaining untransformed volume (also normalised) is plotted on the same graph. The correlation between the variants growth speed and the remaining volume seems fairly good, supporting the proposed interpretation. The underlying physical origin of this behaviour may be related to the presence of a gradient of residual stresses at the surface [20]. The layer close to the free surface will be more affected by the presence of residual compressive stresses opposed to the transformation. The stresses can oppose the propagation of the habit plane in the near surface layer, in a similar manner to the action of oxide layers in the case of martensitic transformation in metals. However, when the transformation propagates (internal growth), the surface effects take more and more importance. Thought still speculative, this effect could explain the observed trend of growth rate. A program is under way in the laboratory to assess more precisely the role of the surface residual stresses on the transformation propagation.

The last trend (stage III) is only observed in the case of needle growth, and corresponds to an "explosive" growth. It has been shown for this mode that transformation strain is not accommodated and stresses consequently building up with the needle growth. These stresses will act to slow down the transformation propagation, their magnitude being so high that the transformation is momentarily stopped.

The energy necessary to complete the transformation was probably brought by an event exterior to the grain. If a neighbouring grain is transforming (as it is the case here), long range stresses will be induced in the surrounding grains. These stresses can overcome the energy barrier and trigger the transformation of the complementary variants all at once, so as to reach a final arrangement where the transformation strains are accommodated on the long range. It is worth noticing that only a *minimum* value of the growth rate in this stage can be provided here. Transformation completion could have occurred at any time during the treatment step. Considering the fact that very large stresses are accumulated, it seems safe assuming the transformation completion occurred all at once (explosive or burst growth), at a speed approaching the sound speed. Stresses are so high that reaching a four-fold configuration with partial transformation (like the one of Fig. 1a) was not favourable at all. The energy disequilibrium is too high to allow partial transformation.

## 3.3 Grain boundary effects

Several transformation-induced relief can be accounted for by the presence and path of grain boundaries, and will be described now.

The first feature is related to the presence of external variants, as described in [17], where secondary external variants are observed on the sides of simple variants arrangement. An example of this is given in Fig. 7, with the corresponding interpretation of the variants arrangement below the free surface. The interesting point to note is the straightforward relationship between the apparent width of variants at the surface and the variants penetration depth below the free surface. Variations of variants width and height at surface will therefore be directly related to a modification of the penetration depth, as shown in Fig. 7. Since the transformation is, at least in the first stages, stopped by the presence of grain boundaries, it can be assumed the penetration depth variations reflects the path of grain boundaries in the volume of the grain, below the free surface. Triggering the transformation on the other side of the grain boundary requires additional stresses, which are not provided in the first stage of the transformation. The transformation induced relief appears thus as directly related to the grains boundaries path and shape below the free surface.

The second effect related to grain boundaries presence have some very important consequences as far as the transformation propagation is concerned, and is related to the formation of microcracks. Fig. 8 provides a case where two parts of the surface (a and b) present a fourfold symmetry, though with a slight disorientation, while the third part (c) presents a very different orientation. The approximate grain orientation, deduced from the fourfold symmetry of surface relief, is represented by the arrows on the first micrograph of Fig. 8.

At the various interfaces of theses zones, microcracks are revealed by AFM observations. This is illustrated in the middle micrograph of Fig. 8, in height mode; the contrast steps reflect the path of microcracks (arrows). If further ageing treatment is performed, microcracking is so extensive that grain pop-out occurs, and some part of the surface are taken away, as shown on the right-hand side micrograph of the figure (arrows), where the remaining holes are easily observed. Holes are located where important microcracking was previously observed, and between grains having the largest disorientation relationships.

These observations allow explaining the whole process of microcracks formation: (a) transformation occurs in two adjacent grains, but having different crystallographic orientations (b) variants grow until the grain boundary is reached (c) microcracks are formed due to transformation accommodation misfit at the grain boundary (d) grain pop-out occurs as a consequence of extensive microcracking.

Finally, if the two adjacent grains present a crystallographic orientation nearly identical, it is possible observing transgranular variants running through the grain boundary. This has already been demonstrated in 3Y-TZP [16], and is also probably the case here in Fig. 9 for CeTZP. The orientation misfit leads to an interruption of the symmetry in the middle of the transformed parts. The estimated grain orientations are given in the figure, along with the grain boundary location (dashed line). The occurrence of low-disorientation grain boundaries is fairly common in ceramics, so that it is a plausible explanation for the relief features observed in Fig. 9.

All these observations can be used to interpret the transformation behaviour at a larger scale, as shown in Fig. 10. A partially transformed zone of larger size is shown in the figure, with various different transformation induced relief features. By using the analysis of the influence of grain boundaries, the presence of microcracks and the approximate orientation derived from the fourfold symmetry, it possible describing the grain boundaries paths (dashed lines). Three grains (1, 4 and 5) presenting a close orientation ($c_t$ axis close to free surface normal) are observed in the middle part. Other grains (2, 3 and 6) with different orientations are observed in the surroundings. Microcracks are observed at the locations of orientation misfit. It is also worth noticing grain pop-out occurred where the misfit was the greatest, i.e. between grains 3 and 4 or 4 and 6. Grains 3 and 6 do not show any evidence of a fourfold symmetry, so that their $c_t$ axis must be away from the free surface normal, and therefore very different of that of grain 4. The misfit is consequently more important than between grains 1 and 4 for which crystallographic orientations are very similar. The formation of microcracks that could later grow in size and potentially lead to critical defects and components failure is here understood at a microscopic scale, in terms of crystallographic arguments. A phenomenon occurring at the scale of the grain has macroscopic consequences. The presence of microcracks in the transformed zone has previously been demonstrated [21, 22], though the underlying critical factors for their formation have never been understood so straightforwardly.

## 4. Discussion

### 4.1 Variants nucleation

As far as the nucleation of the variants is concerned, it is worth resuming the main features of the experimental observations at this point.

- An incubation stage seems to be present during the ageing induced transformation. No relief is observed during the first 60 hrs of the ageing treatment. Since the AFM scans size is obviously limited, it cannot however be ascertained that no variants at all are formed at the surface of the sample during this stage.

- The ABC1 correspondence is the more favourable from an energetic point of view. Grains likely to have their $c_t$ axis close to the surface normal are the first ones to transform [17].
- For this correspondence, three different modes of variants growth have been identified, all of which leading to a final fourfold symmetry. In particular, the partial transformation formed by four opposed variants is a stable configuration, since all the transformation strains are accommodated on long range. In this case, the transformation is proceeding progressively to its completion.
- Variants can propagate into the volume by the external growth mode until an obstacle (i.e. grain boundary) is encountered. Variants then grow from the inside (internal growth). All the variants are thus joining themselves at a grain boundary, at a common origin.
- If isolated single variants are formed (i.e. without a symmetric variant), transformation strains are not accommodated, so that very high level of transformation induced stresses can be obtained locally. Considering the very large energy disequilibrium, the transformation is brought to its completion quasi-instantaneously when stimulated by an external (extra-grain) event. In this case, the variants growth speed is at least larger of one range of order than for stable internal or external growth.

As far as the nucleation of variants leading to the formation of partially transformed grains (Fig 1a) is concerned, two mechanisms can be considered:
- Each one of variants is formed after the other (i.e. needle growth mode), by nucleation at the surface and propagation into the volume as the variant grow. As the magnitude of transformation induced stresses increases with the growth of the variants, these stresses can act to trigger the transformation of neighbouring complementary variants. Hence, it will very unlikely lead to the partial transformation configuration, considering the energy disequilibrium and the above mentioned remarks.
- The four variants composing the final arrangement are formed at the same time, so that no intermediate situation with non-accommodated transformation strains is encountered. The arrangement is stable from its beginning, so that further growth of the variants can proceed slowly and at equilibrium.

Taking into account the partial conclusions presented, the second mechanism is the more favourable candidate able to explain the experimental observations. In this case, the variants must be formed all at once and nucleate at a common origin point. This analysis draws several consequences concerning the chemical mechanism of the transformation, the first one being the nucleation of the transformation in the volume and not at the surface. It was believed the transformation was always starting at the surface of the grains. The analysis proposed here provides evidences of the contrary, at least during the first stages of the transformation and this particular growth mode. When different growth modes are activated (i.e. isolated needle or external growth), the nucleation of the variants occurs at the surface, and variants propagate into the volume as they later grow in size.

From the analysis proposed here, it is possible completing the scenario of the transformation [19, 23, 24]: (a) chemical adsorption of water at the $ZrO_2$ surface (b) reaction of $H_2O$ with $O^{2-}$ on $ZrO_2$ surface to form hydroxyl groups $(OH)^-$ (c) grain boundary diffusion of $(OH)^-$ into the inner part (d) annihilation of the oxygen vacancies by $(OH)^-$ (e) when the oxygen vacancy concentration is reduced so low that the tetragonal phase is no longer stable, t-m transformation occurs at a preferential site and leads to the simultaneous formation of the four opposed variants (f) growth of the opposed variants until the surface is reached. This mechanism could therefore explain both the incubation time and the thermal activation of the transformation observed experimentally, as it would correspond to the time required for the $(OH)^-$ for diffusing along the grain boundary. This is supported by the strong correlation reported between the thickness of the transformed layer and the diffusion distance of $(OH)^-$ [25]. As further action of the water is required for the transformation to propagate, the observed thermal activation of the transformation could correspond to the thermal activation of the $(OH)^-$ diffusion along the grain boundaries. Since the transformation is thermally activated, the thermal treatment steps could have been shortened by electing a higher temperature. Experiments were nonetheless bounded to the technical limits of the autoclave.

## 4.2 Factors affecting the growth stage

Having discussed the various growth mechanisms related to the ABC1 correspondence, the intrinsic origin of the growth mechanism can also be questioned. It has been proven [17] the ABC1 correspondence allows a complete accommodation of transformation strains, so that no residual stresses are induced by the transformation. Continuance of transformation in the surroundings of already transformed parts cannot be accounted by these residual stresses. Two effects can account for the experimental observations:

(1) The first one arises from the classical nucleation theory, which demonstrates the continuation of the transformation is much easier where some parts of the crystal have already transformed (heterogeneous nucleation). In that case, there is no need to create a new interface between the two phases, so that the transformation is much more favourable from an energy point of view.

(2) The second one is related to the approximations made in [17]. The matrix transformation was considerably simplified by the cancellation of the terms having a magnitude smaller than 4/1000. This hypothesis is valid as far as the configuration of transformation induced relief is concerned. However, as nearly all the terms of the transformation matrix are not strictly equal to zero, the ABC1 correspondence does not allow a complete accommodation of the transformation strains. Though of very low magnitude, some residual stresses are expected. These second order stresses can act as a supplementary argument for the continuance of the transformation on previously transformed parts, adding to the first mentioned argument.

The combination of these two effects can rationalize the current experimental observations. If only the favourably orientated grains (with respect to the free surface) are transformed during the first stage, transformation proceeds then by activating less favourable orientation, as shown in Fig. 10. Grains 1, 4 and 5 are the first ones to transform, all of them having their $c_t$ axis close to the free surface normal. Transformation of grains 2, 3 and 6, grains that are more stable, was triggered later, their crystallographic orientation being much less favourable (no fourfold symmetry observed, i.e. the transformation strains accommodation is not as good).

## 4.3 Discussion on the thermodynamics modelling of the transformation

If a tetragonal grain does not transform all at once, as observed experimentally, there should not be a critical grain size below which transformation is not occurring. Moreover, some of the growth mechanisms do not require a nucleation of the transformation in the volume, i.e. diffusion of species is not necessary. In the case of external growth and isolated needle growth, transformation starts at surface (not compulsory at a grain boundary) and then propagates into the volume as the variants grow in size, so that the grain boundaries do not have any importance during the first stages. The growth stage can be affected by a change in grain size, since grain boundaries act as obstacles to the transformation, but the nucleation stage is very little sensitive to the grain size.

It is worth recalling the end point thermodynamic approach of the transformation at this point. The transformation has been analysed in terms of thermodynamic arguments by several authors [7, 8, 26], and the free energy change related to the transformation has been expressed in the following way for a microcrystal of radius r:

$$\Delta G_{q-m} = \tfrac{4}{3}\pi r^3 (\Delta G_{chem} + \Delta G_{dil} + \Delta G_{shear}) + 4\pi r^2 \sum \Delta S_{surface} + A_c \gamma_c V \quad (1)$$

where $\Delta G_{q-m}$ is the total Gibbs energy change, $\Delta G_{chem}$ is the volumic free energy variation, $\Delta G_{dil}$ and $\Delta G_{shr}$ are the free energy variations associated to the dilatational (dil) and shear (shr) components of the transformation, $\sum \Delta S_{surface}$ represents the sum of the terms taking into account the surface effects (surface free energy variation, creation of twin boundaries and creation of interfaces between the t and m phases), and $A_c \gamma_c V$ corresponds to the energy spent in the formation of new surface due to microcracks opening ($A_c$ is the area created by the cracks, $\gamma_c$ the fracture energy per unit area and V the microcrystal volume). Based on the current observations, at least for transformation occurring at the surface, this approach is not very realistic. The present analysis leads indeed to considerable modifications of this equation, in particular when the first stages of the transformation are considered:

- The ABC1 correspondence choice leads to the cancellation of the terms related to the dilatation and shear components of the transformation strains, since all theses transformations strains are accommodated and relaxed by the relief displacement outside of the free surface.
- Transformation does not occur preferentially at the grain boundaries, since the nucleation stage is controlled by the crystallographic orientation of the grains with respect to the free surface. All the created interfaces (junction planes, habit planes) are therefore coherent, at least during the first stages, which does not lead to the formation of additional stresses. All the terms related to surface effects, *but the surface free energy change,* are therefore cancelled.
- No microcrack is formed during the first stage of the transformation.
- Finally, the transformed parts of the volume are not spherical at all, but present a pyramidal configuration. The respective volume and surface area of a partially transformed part can be calculated exactly with calculations outputs.

Let us consider $h_{unt}$ as being the height of the untransformed zone (see fig. 1b), and $h_{tot}$ the height of the transformed zone. The volume $V_{partial}$ of a partially transformed part can be calculated by simple geometric considerations as:

$$V_{partial}(h_{unt}) = 0.416(h_{tot}^3 + h_{unt}^3) \quad (2)$$

The total area of t/m interfaces in the partially transformed arrangement can also be expressed by:

$$A_{partial}(h_{unt}) = 1.385(h_{tot}^2 + h_{unt}^2) \quad (3)$$

Equation (1) becomes thus:

$$\Delta G_{q-m} = 0.416(h_{tot}^3 + h_{unt}^3)\Delta G_{chem} + 1.385(h_{tot}^2 + h_{unt}^2)\Delta S_{chem} \quad (4)$$

The only remaining terms are the chemical free energy variation (in volume and surface). Hence, it is possible quantifying the total free energy change as a function of the variants height, by using values from the literature for the different terms of equation 4, i.e. $\Delta G_{chem}$=-285.10$^6$×(1-T/1448) J.m$^{-3}$ [27] and $\Delta S_{chem}$=0,36 J.m$^{-2}$ [28]. The plot of the total free energy change as a function of the variants height provides the expected critical size (Fig. 11). The value of $\Delta S_{chem}$ was measured in the case of incoherent precipitates, and a critical size for the variants ($h_{tot}$) of 6 nm (for $h_{unt}$=0) is obtained here in that case. Though no values were measured for the case of coherent precipitates, we could expect the surface free energy to decrease of one range of order for the case of coherent precipitates, by comparison with the behaviour of metallic materials. In that case, the critical size falls down to 1 nm (Fig. 11). These values should be compared with the ones obtained when the shear and dilatational components are not accommodated. Since the contribution of these component was very important, the expected critical size (dependent of the temperature), was greater by several ranges of order (typically 1 to 10 µm) [8], and in good agreement with the experimental observations.

By using the outputs of our analysis, it was possible modifying the classical thermodynamic theory to take into account the strain accommodation, the effect of the free surface and the nature of the t-m interfaces. The calculated critical size is much lower than the ones obtained previously, and found so low that no critical size should be expected indeed, as it is not possible manufacturing components with a grain size of a few nanometers. However, it must be kept in mind this approach is reliable only if the shear and dilatational components are accommodated from the very beginning of the transformation. Hypotheses were proposed for the formation of the variants, although no direct observations were possible. If the variants are indeed nucleating within the volume at first, only the formation of the four variants all at once, all of them reaching the surface, would allow this very small critical size. This analysis should however be fairly reliable for variants nucleating at the surface (external growth). For the case of low temperature ageing, the transformation is therefore very different of that occurring during the cooling stage after sintering, where the tetragonal phase can transform spontaneously to its monoclinic structure, and for which the existence of a critical size of grains for transformation have been predicted [7] and demonstrated [29].

## 5. Conclusions

By using atomic force microscopy, the growth mechanisms of tetragonal phase resulting from the martensitic transformation of ceria stabilized zirconia (10 mol% $CeO_2$) are investigated for the first time with a great precision and in a quantitative manner, transformation occurring during low temperature ageing treatments in water vapor. The observations are rationalized by the recent analysis proposed for the crystallographic ABC1 correspondence choice.

- The nucleation is controlled by the crystallographic orientation of the grains. Grains having their $c_t$ axis close the free surface normal are the first ones to transform. Transformation is then propagating by activating different correspondences, less favorable in terms of transformation strains accommodation.
- Three different growth modes have been identified for the ABC1 correspondence: the so-called isolated needle growth and external growth, modes where the transformation starts at surface, and the internal growth, where the transformation is most probably nucleated in the volume and not at the surface.

In that case, an incubation period corresponding to the diffusion time of (OH)⁻ along the grain boundaries should be expected. Three stages were identified for the variants growth kinetics, all of them being related to the degree of transformation strains accommodation.

- Grain boundaries presence and location have a great influence on the transformation induced relief. The penetration depth of the transformation is directly related to the location of grain boundaries during the first stage of the transformation. The transformed relief allows interpreting the grain boundary path straightforwardly. The formation of microcracks, leading to grain pop-out, has been observed at the grain boundary, and is occurring more rapidly between grains having a strong crystallographic disorientation.
- The present experimental observations allow concluding to the absence of the existence of a critical grain size for the transformation, as opposed to the spontaneous transformation during cooling, for which the existence of a critical grain size below which the transformation is not occurring has been previously demonstrated. Moreover, the ageing behavior is essentially similar to that of Y-TZP, which was not clearly established before.

## Acknowledgements

The authors would like to thank the CLAMS for using the nanoscope. Financial support of the European Union under the GROWTH2000 program, project BIOKER, reference GRD2-2000- 25039. Authors are grateful to H. El Attaoui for providing the samples of the study.

| Growth rate (nm/h at 140°C) ±1 | Initial rate (stage I) | Steady rate (stage II) | Burst rate (stage III) |
|---|---|---|---|
| Needle growth | 11 | <3 | >50 |
| Internal growth | 11 | <1 | - |
| External growth | 13 | - | - |

Tab. 1: Growth rate of martensite variants for the different growth modes.

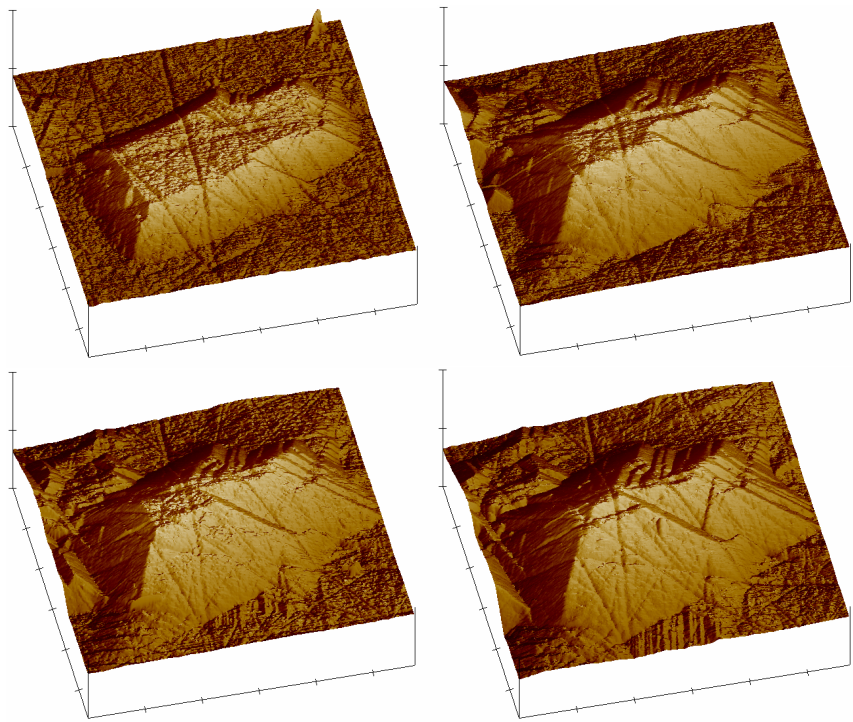

Fig. 1a: Observation of partial transformation and internal growth. Ageing steps of 20 hrs. Horizontal scale: 1 µm/div, vertical scale: 250 nm/div.

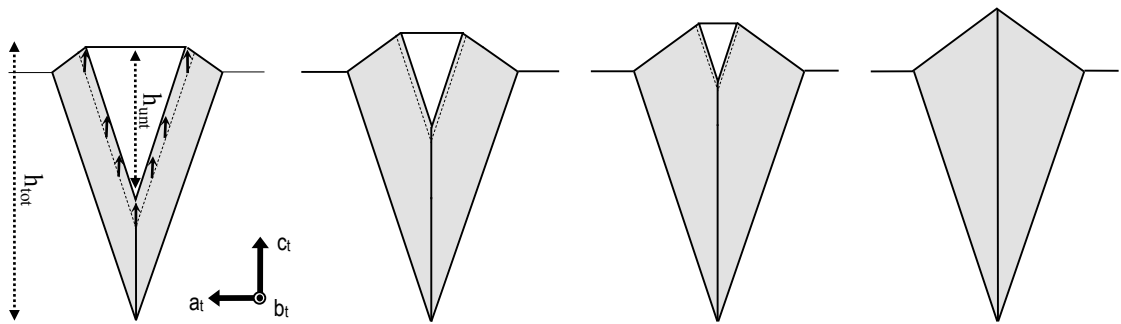

Fig. 1b: Interpretation of the arrangement observed in fig. 1a. The vertical scale is not respected for clarity. The dashed lines represent the location of the inner habit planes before transformation.

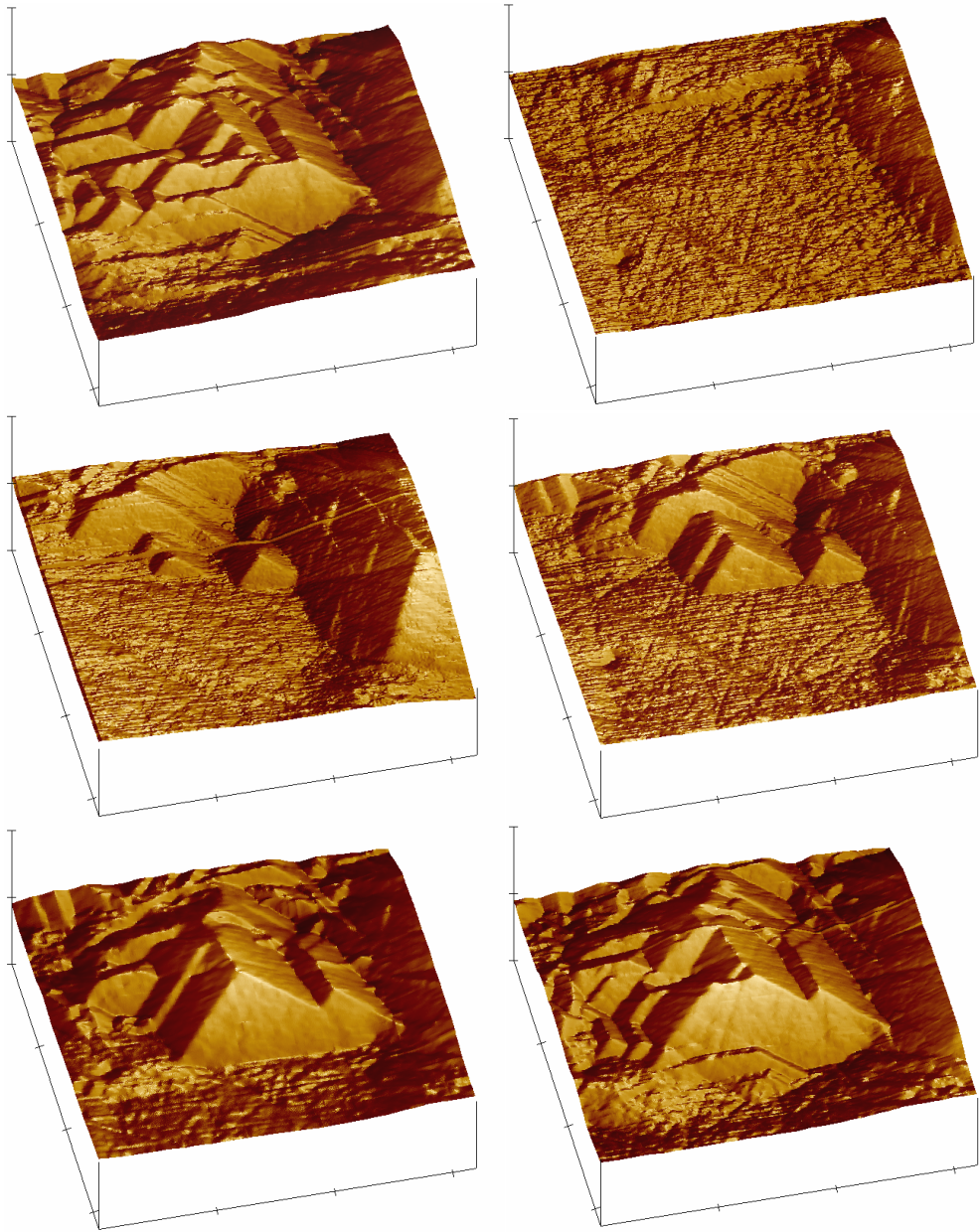

Fig. 2a: Observation of external growth. Ageing steps of 20 hrs.

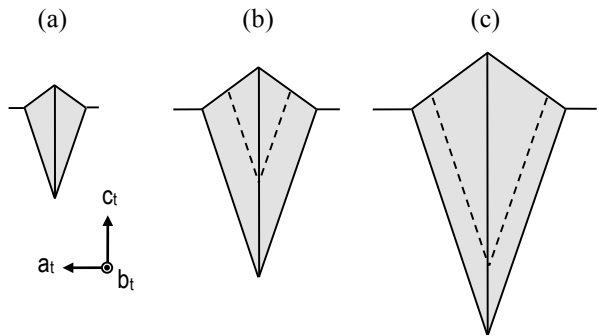

Fig. 2b: Interpretation of the arrangement observed in fig. 2a. The position of the junction plane remains constant along the transformation. The vertical scale is not respected for clarity. The dashed lines represent the location of the habit planes at the previous transformation stage.

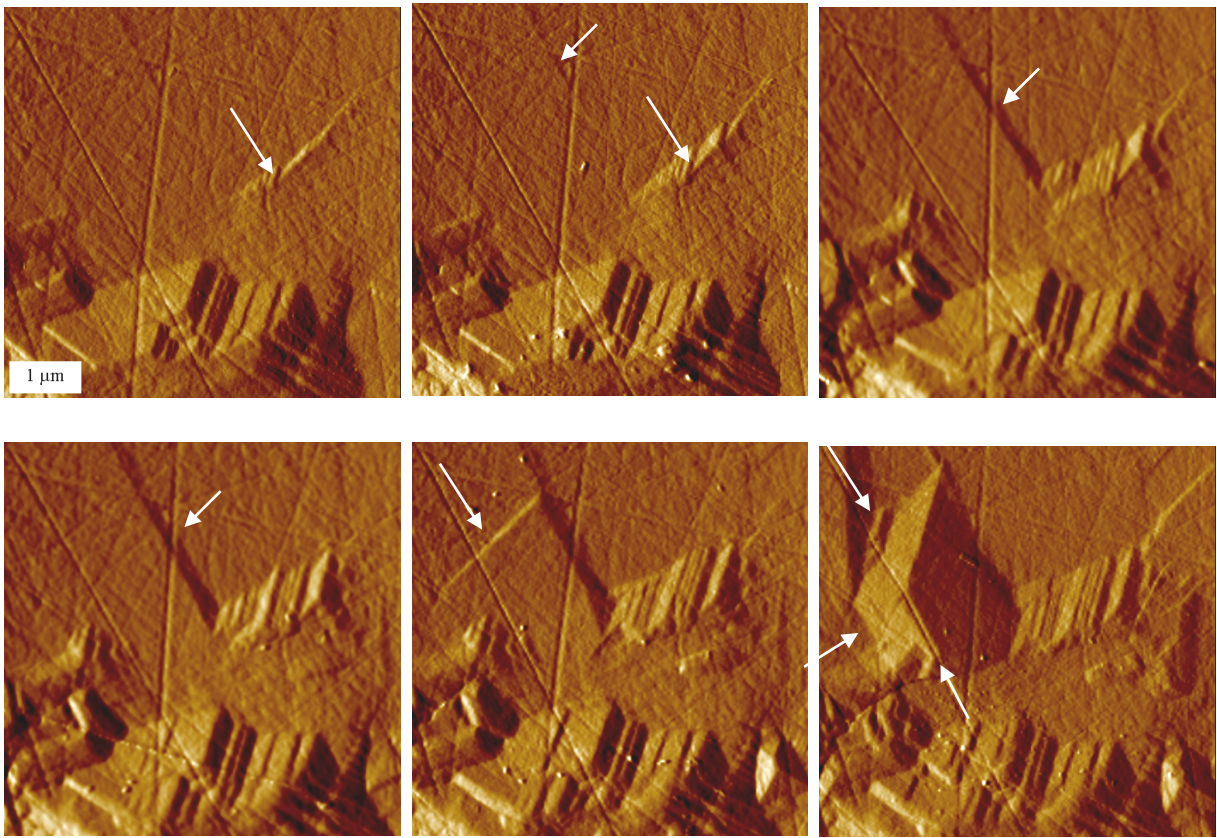

Fig. 3: Observation of consecutive isolated needles growth. Ageing steps of 20 hrs.

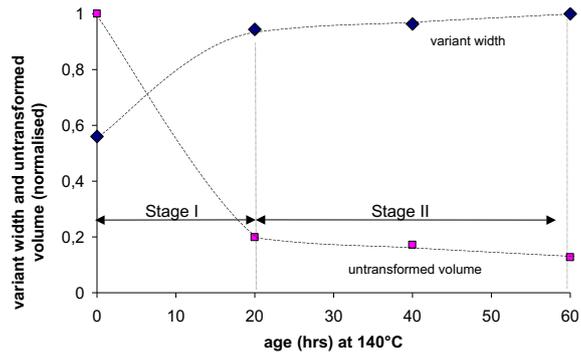

Fig. 4: Normalised variant width as a function of its age, internal growth mode.

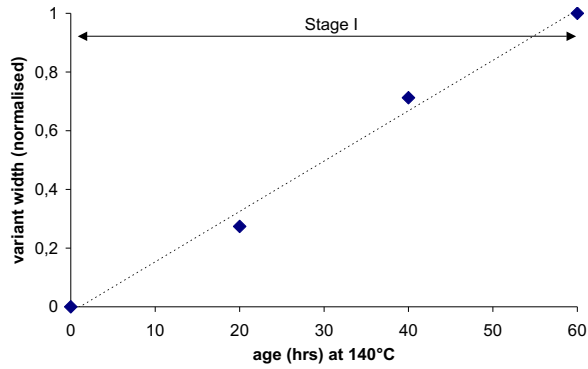

Fig. 5: Normalised variant width as a function of its age, external growth mode.

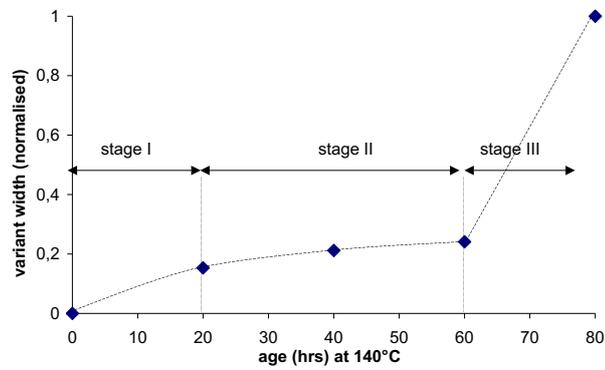

Fig. 6: Normalised variant width as a function of its age, isolated needle growth mode.

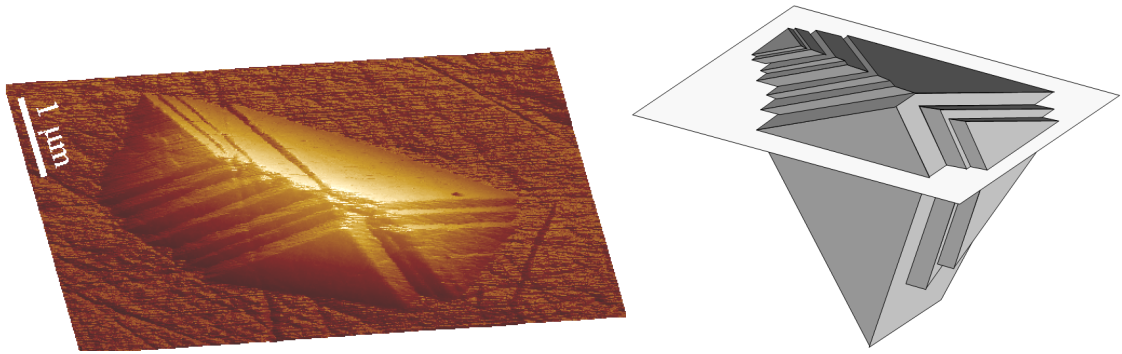

Fig. 7: Observation and interpretation of more complex relief features with secondary external variants. The decrease of the width and height at surface is related to a decrease of the penetration depth of the variants.

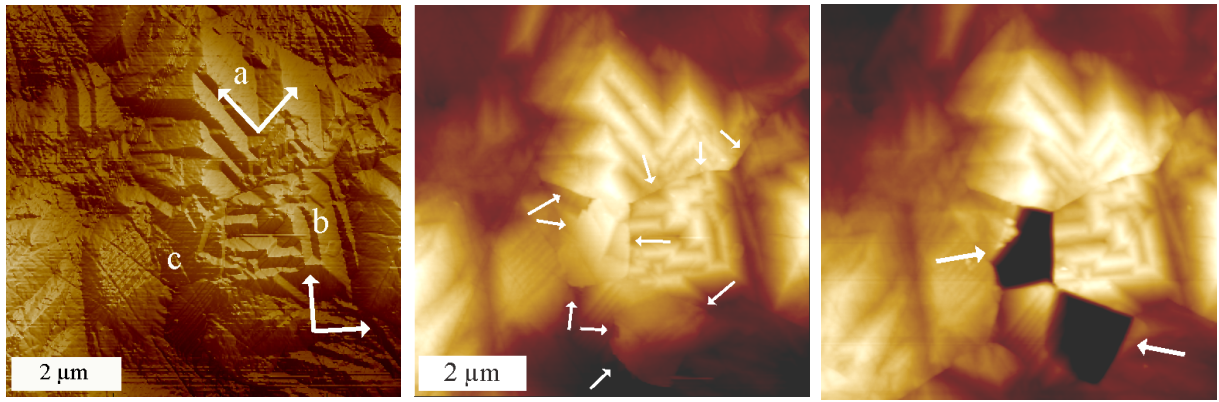

Fig. 8: Formation of microcracks at the grain boundaries between three grains (denoted *a*, *b* and *c*), 3D and height image (left and middle) and grain pop-out (arrows), 20 hrs later (right, height image). The arrows on the first micrograph represent the approximate orientation of the $a_t$ and $b_t$ axis of the two adjacent grains. Arrows on the other micrographs indicate the microcracks location. Grains *a* and *b* have both the $c_t$ axis close the free surface normal, while grain *c* is differently oriented.

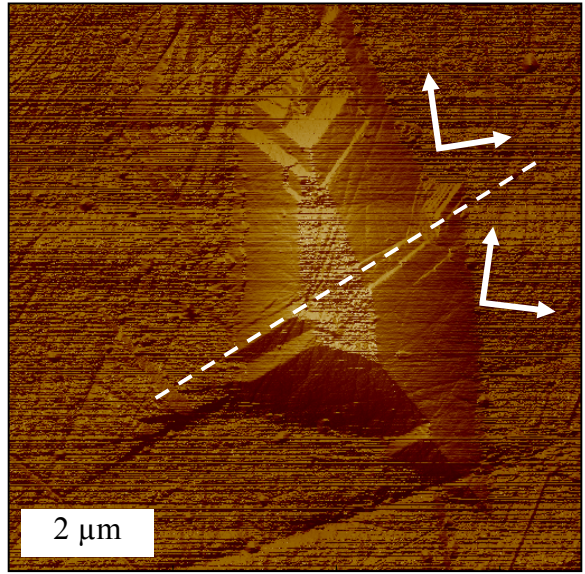

Fig. 9: Probable grain boundary of low disorientation, leading to the observed disrupted relief of transgranular variants. The approximate orientation of the $a_t$ and $b_t$ axis of the two adjacent grains is plotted on the micrograph, along with the position of the grain boundary (dashed line)

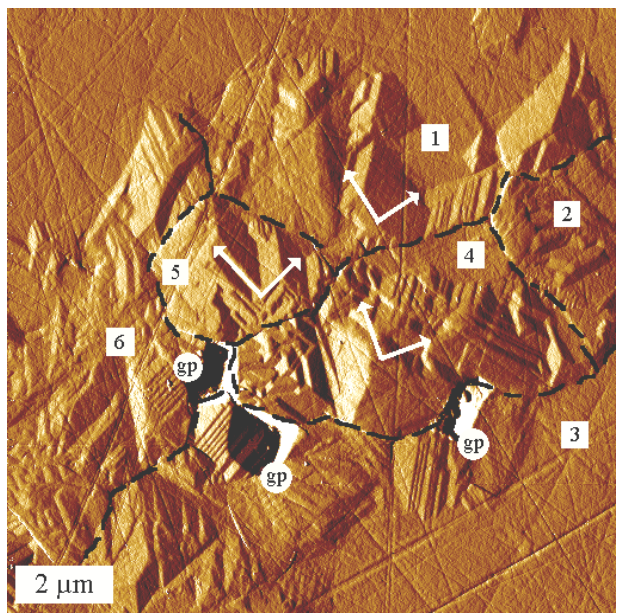

Fig. 10: Observation of transformation induced relief in a large portion of the surface. The relief features allow the approximate determination of the location of grain boundaries. Deduced grains are numbered, and grain pop-out (gp) locations are indicated. Grains 1, 4 and 5 present a very low crystallographic disorientation. Dashed lines represent the expected grain boundaries.

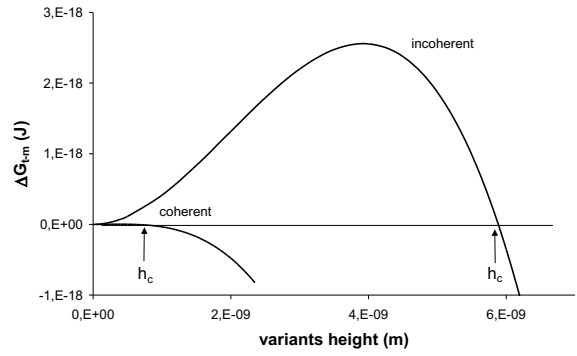

Fig. 11: Total free energy change of a stack of four accommodated variants as a function of the variants height. The change in energy was estimated for the case of coherent and incoherent t-m interface. The critical size for transformation is much smaller in the case of a coherent interface.